\documentstyle[eqsecnum,prd,aps]{revtex}

\begin{document}

\draft
\title{SOLVING THE HAMILTON-JACOBI EQUATION FOR \\
GENERAL RELATIVITY}

\author{J. Parry,$^1$ D. S. Salopek$^{1,2}$ and J. M. Stewart$^1$}
\vspace{2pc}
\address{
{}$^1$University of Cambridge,
Department of Applied Mathematics and Theoretical Physics \\
Silver Street, Cambridge, CB3 9EW, UK }
\vspace{2pc}
\address{
{}$^2$Department of Physics, University of Alberta,
Edmonton, Canada T6G 2J1 }

\date{\today}

\maketitle

\begin{abstract}
We demonstrate a systematic method for solving the Hamilton-Jacobi
equation for general relativity with the inclusion of matter fields.
The generating functional is expanded in a series of spatial gradients.
Each term is manifestly invariant under reparameterizations of the
spatial coordinates (``gauge-invariant'').  At each order we solve
the Hamiltonian constraint using a conformal transformation of the
3-metric as well as a line integral in superspace.
This gives a recursion relation for the generating functional which
then may be solved to
arbitrary order simply by functionally differentiating previous orders.
At fourth order in spatial gradients, we demonstrate  solutions
for irrotational dust as well as for a scalar field.
We explicitly evolve the 3-metric  to the same order.
This method can be used to derive the Zel'dovich approximation
for general relativity.

\end{abstract}


%\tighten  % to get single-spaced preprint
%\twocolumn

%\narrowtext
%\mediumtext
\widetext

\section{Introduction}

In applications to cosmology, Hamilton-Jacobi (HJ) theory plays
a vital role in describing the gravitational field.  It serves
as a starting point for a semi-classical analysis used both in
stochastic inflation \cite{Starobinsky}, \cite{uslot}, \cite{SB},
and in quantum cosmology \cite{Hall.Hawk}.
Recently it has been shown how to apply HJ techniques in solving
the classical Einstein's equations.  For example, one may derive
the Zel'dovich approximation describing the formation of pancake
structures in a matter-dominated Universe \cite{Zel}, \cite{CPSS}.

The Hamilton-Jacobi equation for pure gravity was first derived
by Peres \cite{Peres}. He showed that the momentum constraint
implies the generating functional or the phase of the semi-classical
wavefunctional remains invariant under arbitrary reparameterizations
of the spatial coordinates (diffeomorphism invariance).
However he was not able to solve the HJ equation.  We find that a
crucial ingredient is the introduction of matter fields.
We also capitalize on the diffeomorphism invariance of the generating
 functional.

In this paper we expand the generating functional in a series of
spatial gradients. This approximation scheme has a long history
dating back to Lifshitz and Khalatnikov \cite{L.and.K}, as
well as Eardley, Liang and Sachs \cite{ELS}.
In a quantum context, Teitelboim \cite{Teitelboim} and
Pilati \cite{Pilati} considered an analogous approach which
they called the  strong-coupling expansion.
Accurate to first order in spatial gradients, Salopek and Bond
\cite{SB} explicitly solved the Hamilton-Jacobi equation and
the momentum constraint with scalar fields present.  In an important
generalization, Salopek and Stewart \cite{S.and.S}
added  dust fields and then demonstrated
solutions to second order.  The principal goal of this paper is
to give solutions to higher order.

New techniques yield a practical and systematic method of computing
higher order terms.
It proves convenient to employ a conformal transformation of the 3-metric
to simplify the Hamiltonian constraint, which can then be integrated
using a line integral in superspace.
A comparison with an exact solution shows that
in situations of interest for cosmology even the first few
terms can be quite accurate.

We start in Sec. {\uppercase\expandafter{\romannumeral 2}} by writing
out the Hamilton-Jacobi equation for general relativity including
two types of matter fields, a regular scalar field $\phi$, and a
dust field $\chi$ describing a pressure-free fluid of irrotational
massive particles, for example, cold-dark-matter.
The spatial gradient expansion is introduced, and the Hamilton-Jacobi
equation is expanded to all orders in spatial gradients.
Solutions of this equation correct to zero and two spatial gradients are
 reviewed.
A new technique involving a conformal transformation is introduced
in Sec. {\uppercase\expandafter{\romannumeral 3}} in order to
find solutions to higher orders.  A recursion
relation is derived which enables one to compute
solutions to any
given order, and we explicitly give the fourth order terms.
For a single dust field, we solve the evolution equation for the 3-metric in
Sec.  {\uppercase\expandafter{\romannumeral 4}}, and  show that
our solution recovers the Szekeres exact solution \cite{SZ} to Einstein's
equations.
This is followed by some concluding remarks.
(Units are chosen so that $c=8\pi G= \hbar= 1$.
The  sign conventions of Misner, Thorne and
Wheeler \cite{MTW} will be adopted throughout.)

\section{The Hamilton-Jacobi equation for general relativity}

In HJ theory the primary object of interest is the generating functional
$\cal S$, which is the phase of the semi-classical wavefunctional
$\Psi\sim e^{i{\cal S}}$.
For general relativity, the HJ equation is simply the energy constraint with
 the momenta replaced by functional derivatives of the 3-metric.
In addition we must consider the
momentum constraint, which demands the gauge invariance of $\cal S$.

In this section we write $\cal S$ as a series of terms grouped
according to the number of spatial derivatives that they contain.
The Hamilton-Jacobi equation is likewise written as a sum of such terms.
  We review solutions for $\cal S$ correct to zeroth order
(no spatial gradients) and to second order (two spatial gradients) for a
 single dust field, and for a single scalar field.

The action for general relativity interacting with a scalar field $\phi$
 and a dust field $\chi$ is
\begin{equation}
{\cal I }=\int d^4 x \sqrt{-g} \left( {1\over 2}{\hspace{-0.1cm}}\ ^{(4)}
{\hspace{-0.04cm}}R
- {1\over 2} g^{\mu\nu}\partial_{\mu}\phi
\partial_{\nu}\phi
-V(\phi , \chi )  - {1\over 2m} n\left( g^{\mu\nu}\partial_{\mu}\chi
\partial_{\nu}\chi + m^2\right)
\right),
\label{fullaction}
\end{equation}
where$\ ^{(4)}R$ is the Ricci scalar of the space-time metric $g_{\mu\nu}$,
 $\phi$ is a scalar field and $\chi$ is a velocity potential for irrotational
 dust particles of rest mass $m$ (which is a universal constant).
 $V(\phi,\chi)$ is a potential for the matter fields.
The rest number density $n\equiv n(t,x)$ is a Lagrange multiplier which
 ensures that the 4-velocity
\begin{equation}
U^\mu = -g^{\mu\nu}\chi_{,\nu}/m
\label{4velocity}
\end{equation}
satisfies $U^\mu U_\mu = -1$.
$\chi$ may be rescaled to include $m$, and so from now on we set $m=1$.

In the ADM formalism the line element is written
\begin{equation}
ds^2=\left(-N^2+\gamma^{ij}N_iN_j\right)dt^2 + 2N_idt\,dx^i
+ \gamma_{ij}dx^i\,dx^j\ ,
\label{ADMdecomp}
\end{equation}
where $N$ and $N_i$ are the lapse and shift functions respectively,
and $\gamma_{ij}$ is the 3-metric.  We can then rewrite the action
in Hamiltonian form,
\begin{equation}
{\cal I}=\int d^4x\left(\pi^{\Phi_a}\dot\Phi_a +\pi^{ij}\dot\gamma_{ij}
-N{\cal H} -N^i{\cal H}_i\right).
\label{ADMaction}
\end{equation}
Here the $\pi^{ij}$ denote the momenta conjugate to $\gamma_{ij}$
and $\Phi_a =(\phi,\chi)$ is an assembly of the scalar and dust fields
 with corresponding conjugate momenta $\pi^{\Phi_a}$.  We use the summation
 convention over the subscript $a=1,2$.  Variations of the lapse and shift
 functions produce the Hamiltonian and momentum constraints:
\begin{mathletters}
\label{constraints}
\begin{eqnarray}
{\cal H}=&&\gamma^{-1/2}\pi^{ij}\pi^{kl}
\left[2\gamma_{il} \gamma_{jk} - \gamma_{ij}\gamma_{kl}\right]
+\sqrt{1 + \gamma^{ij}\chi_{,i}\chi_{,j}}\,\pi^\chi
+ {1\over 2} \gamma^{-1/2}\left(\pi^\phi\right)^2 +\gamma^{1/2}V(\Phi_a)
\nonumber \\
&& -{1\over 2}\gamma^{1/2}R
+{1\over 2} \gamma^{1/2}\gamma^{ij}\phi_{,i}\phi_{,j} = 0\ ,
\label{hamconstraint} \\
{\cal H}_{i}=&&-2\left(\gamma_{ik}\pi^{kj}\right)_{,j}
 + \pi^{lk}\gamma_{lk,i} +
 \pi^{\Phi_a} \Phi_{a,i} = 0 \ ,
\label{momconstraint}
\end{eqnarray}
\end{mathletters}
where $R$ is the Ricci scalar of the 3-metric.

Variation of the action (\ref{ADMaction}) with respect to the momenta give
 the evolution equations for the field variables
\begin{mathletters}
\label{evolutioneqns}
\begin{equation}
\left(\dot\phi-N^i\phi_{,i}\right)/N=\gamma^{-1/2}\pi^\phi,
\label{evol.scalar}
\end{equation}
\begin{equation}
\left(\dot\chi-N^i\chi_{,i}\right)/N=\sqrt{1+\chi_{,i}\chi^{,i}},
\label{evol.dust}
\end{equation}
\begin{equation}
\left(\dot\gamma_{ij}-N_{i|j}-N_{j|i}\right)/N =2\gamma^{-1/2}\pi^{kl}
\left(2\gamma_{jk}\gamma_{il}-\gamma_{ij}\gamma_{kl}\right),
\label{evol.metric}
\end{equation}
\end{mathletters}
where $|$ denotes covariant differentiation with respect to the 3-metric.
Variations of the action (\ref{ADMaction}) with respect to the field variables
 yield the evolution equations for the momenta.
However, these remaining evolution equations  are automatically
satisfied provided that
\begin{equation}
\pi^{ij}(x)={\delta{\cal S}\over \delta{\gamma_{ij}(x)}}\ , \qquad
\pi^{\Phi_a}(x)={\delta{\cal S}\over \delta\Phi_a (x)}\ ,
\label{pis}
\end{equation}
satisfy the constraint equations (\ref{constraints})
and provided the evolution equations
(\ref{evolutioneqns}) hold.
In this interpretation the energy constraint (\ref{hamconstraint}) is a
self-contained equation for $\cal S$, whereas the momentum constraint
(\ref{momconstraint}) demands that $\cal S$ be gauge-invariant.  The
lapse and shift functions do not enter the constraints.
Loosely speaking, equations (\ref{evolutioneqns})  imply that trajectories
in superspace are orthogonal to the phase functional.
(In a more general formulation, $\cal S$ actually generates a canonical
transformation to new variables \cite{S.and.S}.)

In terms of the generating functional,
the Hamiltonian and momentum constraints are
\begin{mathletters}
\label{Sconstraints}
\begin{eqnarray}
{\cal H}(x)=&&\gamma^{-1/2} {\delta{\cal S}\over \delta\gamma_{ij}(x)}
{\delta{\cal S}\over \delta\gamma_{kl}(x)}
\left[2\gamma_{il}(x) \gamma_{jk}(x) - \gamma_{ij}(x)\gamma_{kl}(x)\right]
\nonumber \\
&& + {1\over 2} \gamma^{-1/2}\left({\delta{\cal S}\over \delta\phi(x)}
\right)^2 +
 \sqrt{1 + \gamma^{ij}\chi_{,i}\chi_{,j}}\,
{\delta{\cal S}\over\delta\chi (x)}+\gamma^{1/2}V(\Phi_a(x)) \nonumber \\
&& -{1\over 2}\gamma^{1/2}R
+{1\over 2} \gamma^{1/2}\gamma^{ij}\phi_{,i}\phi_{,j}=0\ ,
\label{HJequation} \\
{\cal H}_{i}(x)=&&-2\left(\gamma_{ik}{\delta{\cal S}\over
\delta\gamma_{kj}(x)}
\right)_{,j} +
{\delta{\cal S}\over\delta\gamma_{lk}(x)}\gamma_{lk,i} +
 {\delta{\cal S}\over\delta\Phi_a (x)} \Phi_{a,i}=0\ .
\label{Smomentum}
\end{eqnarray}
\end{mathletters}
Equation (\ref{HJequation}) is now called the Hamilton-Jacobi equation.

\subsection{Spatial Gradient Expansion of the Generating Functional}

Our method will be to expand the generating functional
\begin{equation}
{\cal S}= {\cal S}^{(0)} + {\cal S}^{(2)} + {\cal S}^{(4)} + \dots\ ,
\label{theexpansion}
\end{equation}
in a series of terms according to
the number of spatial gradients they contain.
As a result the Hamilton-Jacobi equation can be
grouped into terms with an even number of spatial
derivatives:
\begin{equation}
{\cal H}={\cal H}^{(0)} + {\cal H}^{(2)} + {\cal H}^{(4)} + \dots\ .
\label{theexpansion2}
\end{equation}

\subsubsection{Solution of zeroth order Hamiltonian}

The Hamilton-Jacobi equation of order zero is
\begin{eqnarray}
{\cal H}^{(0)}=&&
\gamma^{-1/2} {\delta {\cal S}\over \delta\gamma_{ij}}^{(0)}
{\delta{\cal S}\over \delta\gamma_{kl}}^{(0)}
\left(2\gamma_{jk}\gamma_{li}- \gamma_{ij}\gamma_{kl}\right)
+{1\over 2} \gamma^{-1/2}\left(
{\delta{\cal S}\over \delta\phi}^{(0)}\right)^2
\nonumber \\
 && + {\delta{\cal S}\over \delta\chi}^{(0)}
 + \gamma^{1/2}V\left(\Phi_a\right) =0.
\label{zeroham}
\end{eqnarray}
The diffeomorphism invariance of the generating functional
suggests a solution of the form
\begin{equation}
{\cal S}^{(0)}=-2\int d^3x\,\gamma^{1/2}\, H\left(\Phi_a\right).
\label{zeroth}
\end{equation}
Because the integral is over $d^3x\,\gamma^{1/2}$  the functional is
invariant
 under coordinate transformations, and it clearly contains no
spatial derivatives.
The numerical factor $-2$ is chosen so that $H$ corresponds to
the usual Hubble parameter in the long-wavelength approximation.
The following condition must be satisfied in order that
the zeroth energy constraint vanish:
\begin{equation}
H^2={2\over 3}\left({\partial H\over\partial\phi}\right)^2
    -{2\over3} {\partial{H}\over\partial\chi}
   +{1 \over 3}V\left(\phi,\chi\right)\ .
\label{Hequation}
\end{equation}
We now consider two solutions of this nonlinear equation.
Firstly for a single dust field and vanishing potential it is trivial
to show that
\begin{equation}
H(\chi)={2\over 3\chi}\ ,
\label{zero.dust}
\end{equation}
is a solution.
Setting the shift function to zero, and choosing our time
coordinate to be $\chi$ (i.e., the lapse function $N=1$),
solution of the metric evolution equation (\ref{evol.metric}) gives
\begin{equation}
\gamma^{(1)}_{ij}(\chi,x)=\chi^{4/3} \ k_{ij}(x)
\label{first}
\end{equation}
where $k_{ij}(x)$, the `seed' metric, is an arbitrary function
of the spatial variables.
This is the standard result for a dust-dominated universe
at long-wavelengths.

Secondly we consider a scalar field self-interacting with an
exponential potential
\begin{equation}
V(\phi)=V_0\exp\left(-\sqrt{2\over p}\phi\right),
\label{s.f.potential}
\end{equation}
where $p$ is a constant.
For this case, one can find analytic solutions.
The equation (\ref{Hequation}) has an attractor solution
\begin{equation}
H(\phi)=\left[{V_0\over 3(1-1/3p)}\right]^{1/2}
\exp\left(-{\phi\over\sqrt{2p}}\right).
\label{zero.scalar}
\end{equation}
For more details see ref. \cite{SB}.

\subsubsection{Solution of the second order Hamiltonian}

In this section we will give a brief review of the second order,
which was treated in Salopek and Stewart \cite{S.and.S}.

The second order HJ equation is
\begin{eqnarray}
{\cal H}^{(2)}=&&2H\gamma_{ij}
{\delta{\cal S}^{(2)}\over\delta\gamma_{ij}}
   -2{\partial{H}\over\partial\phi}{\delta{\cal S}^{(2)}\over\delta\phi}
        +{\delta{\cal S}^{(2)}\over \delta\chi}
\nonumber \\
  &&+ {1\over 2}\gamma^{ij}\chi_{,i}\chi_{,j}
     {\delta{\cal S}^{(0)}\over\delta\chi}
    - {1\over 2}\gamma^{1/2}R
    + {1\over 2}\gamma^{1/2}\gamma^{ij}\phi_{,i}\phi_{,j}=0.
\label{secondham}
\end{eqnarray}
Their method was to write an Ansatz for the functional ${\cal S}^{(2)}$
which contains all terms that are second order in spatial gradients and
are diffeomorphism invariant.  There are three types of terms ---
$R, \Phi_{a,i}\Phi^{a,i}$ and $\Delta\Phi_a$.
\begin{equation}
{\cal S}^{(2)}=\int \gamma^{1/2}d^3x
\left(J(\Phi_a)R + K^{ab}(\Phi_c)\Phi_{a|i}\Phi_b^{\ |i}
+L^a(\Phi_b)\Delta\Phi_a
\right).
\label{second.ansatz}
\end{equation}
The last term may be integrated by parts, and by redefining
\begin{equation}
K^{ab}\rightarrow K^{ab}- {\partial{L}^a\over\partial\Phi_b},
\label{redef.k}
\end{equation}
we may set $L^a=0$.  Substitution of the
Ansatz (\ref{second.ansatz}) into the Hamilton-Jacobi equation
at this order (\ref{secondham})
yields three sets of coupled partial differential equations
which then must be solved.
The number of equations exceeds the number of unknowns.  However,
no contradiction arises because some are redundant for reasons
which will be discussed in the next section.

For a single dust field there are two independent equations.
\begin{mathletters}
\label{2nd.dust.direct}
\begin{equation}
HJ+{\partial{J}\over\partial\chi}={1/2}\ ,
\end{equation}
\begin{equation}
2H{\partial J\over\partial\chi} + K=0\ .
\end{equation}
\end{mathletters}
Given the zeroth order solution (\ref{zero.dust}), these can be solved
to give
\begin{equation}
{\cal S}^{(2)}=\int d^3x\,\gamma^{1/2}\,\left\{
\left({3\over10}\chi + D \chi^{-2/3}\right) R
 +\left(-{2\over5}\chi^{-1}+{8\over9} D \chi^{-8/3}\right)
\gamma^{ij}\chi_{,i}\chi_{,j}
\right\}\ .
\label{second.fluid}
\end{equation}
The term proportional to the arbitrary integration constant $D$
will be refered to as the complementary functional.

In an analogous way, for a single scalar field with the zeroth order
solution (\ref{zero.scalar}), we find (neglecting the complementary
functional)
\begin{equation}
{\cal S}^{(2)} = \int d^3x\,\gamma^{1/2}
{p\over 2(p+1)}{1\over H}\left(R - \gamma^{ij}\phi_{,i}\phi_{,j}\right).
\label{second.scalar}
\end{equation}

\section{SOLVING THE HIGHER ORDER HAMILTONIAN}

The Ansatz method that was used to solve the second order case
can be generalized to higher orders, but it is extremely tedious
because the number of terms increases dramatically.  For example the
fourth order case has 15 types of
diffeomorphism invariant terms leading to 15 coupled sets of partial
differential equations.  Although some of the free functions can be
set to zero, just as in Sec. \uppercase\expandafter{\romannumeral 2}.B.2,
there are still  eight variables satisfying fifteen equations.
It is difficult to proceed further along these lines,
and so we adopt an alternative approach.

Using the expression for ${\cal S}^{(0)}$, (\ref{zeroth}),
we write down the Hamiltonian of order $2n$:
\begin{mathletters}
\label{Htwon}
\begin{equation}
{\cal H}^{(2n)}=
 \left\{ 2H\gamma_{ij}{\delta{\cal S}^{(2n)}\over\delta\gamma_{ij}}
-2 {\partial{H}\over\partial\phi}
{\delta{\cal S}^{(2n)}\over\delta\phi}
+ {\delta{\cal S}^{(2n)}\over\delta\chi} \right\}
+{\cal R}^{(2n)} =0\ .
\label{Htwona}
\end{equation}
Here the remainder term ${\cal R}^{(2n)}$, which is independent of
${\cal S}^{(2n)}$, contains contributions
from all previous orders, and is assumed to be known:
\begin{eqnarray}
 {\cal R}^{(2n)} =&&
  \gamma^{-1/2}\sum_{p=1}^{n-1}
 {\delta{\cal S}^{(2p)}\over\delta\gamma_{ij}}
 {\delta{\cal S}^{(2n-2p)}\over\delta\gamma_{kl}}
 \left(2\gamma_{jk}\gamma_{li}- \gamma_{ij}\gamma_{kl}\right)
+ \gamma^{-1/2}\sum_{p=1}^{n-1}{1\over 2} {\delta{\cal S}^{(2p)}
\over\delta\phi}
    {\delta{\cal S}^{(2n-2p)}\over\delta\phi}
\nonumber \\
&&+\sum_{p=1}^n  {{\scriptstyle 1\over2}\choose \scriptstyle p}
 \left(\gamma^{ij}\chi_{,i}\chi_{,j}\right)^p
{\delta{\cal S}^{(2n-2p)}\over\delta\chi} + {\cal V}^{(2n)}\ ,
\label{def.of.Q}
\end{eqnarray}
where the superspace potential ${\cal V}^{(2n)}$ is defined to be
\begin{equation}
{\cal V}^{(2n)} =
\left\{
\begin{array}{ll}
-{1\over 2}\gamma^{1/2}R +
{1\over 2}\gamma^{1/2}\gamma^{ij}\phi_{,i}\phi_{,j}
&  \ \mbox{for $n=1$,} \\
 0 & \ \mbox{otherwise.}
\end{array}
\right.
\end{equation}
\end{mathletters}Although the Hamiltonian constraint of order $2n$,
(\ref{Htwon}), is
basically a
linear equation for ${\cal S}^{(2n)}$ (unlike the zeroth order
equation (\ref{zeroham})), it is nevertheless not immediately
apparent how to calculate ${\cal S}^{(2n)}$ given the previous
orders, partly because the term in braces in equation
(\ref{Htwona}) is complicated.  In this section we will show,
firstly for a
single dust field, and then for a single scalar field, a technique
which yields a recursion relation for the generating functional.
Finally we indicate briefly how to extend the analysis to multiple
fields.

\subsection{Single Dust Field}

To simplify the HJ equation (\ref{Htwon}) for a single dust field,
we utilise a conformal transformation of the 3-metric to
define a metric variable
\begin{equation}
f_{ij}(x) \equiv\Omega^{-2}(\chi(x)) \  \gamma_{ij}( x)
\label{conformal.trans.dust}
\end{equation}
where $\Omega$ satisfies
\begin{equation}
{\partial\Omega\over \partial\chi} = H\Omega\ .
\label{eqn.for.Omega.dust}
\end{equation}
The HJ equation (\ref{Htwona}) reduces to
\begin{equation}
\left. {\delta {\cal S}^{(2n)}\over\delta\chi}\right|_{f_{ij}}
+ {\cal R}^{(2n)}
=0\ .
\label{HJ.conf.for.dust}
\end{equation}
We note immediately that there are some obvious solutions
to the homogeneous version of this equation
$\left.\delta{\cal S}^{(2n)}/\delta\chi\right|_{f_{ij}}=0$
which correspond to the complementary functionals.  Any functional
of only the conformal metric $f_{ij}$ that has $(2n)$ spatial
gradients and is diffeomorphism invariant will satisfy this equation.
For example, at the
second order ${\cal S}^{(2)}= D \int d^3x f^{1/2} \ \tilde R$ is the
complementary functional, where $\tilde R$ is the Ricci scalar of
$f_{ij}$.  This corresponds to the arbitrary part of
(\ref{second.fluid}).

\subsubsection{The second order revisited}

To show how to use this conformal transformation, we shall
now return to the second order HJ equation with one dust
field present,
\begin{equation}
\left. {\delta {\cal S}^{(2)}\over\delta\chi}\right|_{f_{ij}}
 = -{\cal R}^{(2)}
={1\over 2}\gamma^{1/2}R
+\gamma^{1/2} {\partial H\over\partial\chi}\gamma^{ij}\chi_{,i}\chi_{,j}\ .
\end{equation}
The right hand side of this may be expressed in terms of the conformal
metric $f_{ij}$ using the standard rules for canonical transformations,
for example,
\begin{equation}
R^i_{\ j}=\Omega^{-2}\tilde R^i_{\ j} +
\Omega^{-1}\left(\Omega^{-1}\right)_{;jk}f^{ik}
- \Omega^{-3}f^{kl}\Omega_{;kl}\delta^i_{\ j}\ ,
\end{equation}
where $R^i_{\ j}$ is the Ricci tensor of the 3-metric $\gamma_{ij}$,
$\tilde R^i_{\ j}$ is the Ricci tensor of $f_{ij}$, and a semicolon
denotes covariant differentiation with respect to $f_{ij}$.
Rewriting the right hand side and using the definition of the
conformal factor (\ref{eqn.for.Omega.dust}), we find
\begin{equation}
\left. {\delta {\cal S}^{(2)}\over\delta\chi(x)}\right|_{f_{ij}} =
-{\cal R}^{(2)}\left[\chi(x), f_{ij}(x)\right]=
f^{1/2}\left( {1\over2}\Omega \tilde R
-f^{ij}\chi_{,i}\chi_{,j} {\partial\ \over\partial\chi}(H\Omega)
- 2 f^{ij}\chi_{;ij} (H\Omega)\right) \ .
\label{2nd.revisit.diff}
\end{equation}
By inspection the solution is
\begin{mathletters}
\label{second.order.in.f}
\begin{equation}
{\cal S}^{(2)}=\int d^3x f^{1/2} \left[ j(\chi)\tilde R
+ k(\chi)\chi^{;i}\chi_{;i}\right]\ ,
\label{second.form.of.S}
\end{equation}
where the $\chi$-dependent coefficients $j$ and $k$ are
\begin{equation}
j(\chi)=\int_0^\chi {\Omega(\chi')\over 2}\;d\chi'\ + D ,\qquad
k(\chi)=H\Omega\ ,
\label{j.and.k.for.dust}
\end{equation}\end{mathletters}and
$D$ is a constant. We can use the conformal transformation
rules to find the generating functional in terms of the 3-metric
$\gamma_{ij}$
by using the zeroth order solution (\ref{zero.dust}) and
the conformal factor
\begin{equation}
\Omega\equiv \chi^{2/3}\ ,
\label{Omega.for.dust}
\end{equation}
which satisfies (\ref{eqn.for.Omega.dust}).
This procedure recovers the generating functional (\ref{second.fluid})
obtained by the
direct approach.  The constant of integration $D$ in the
definition of $j$ yields  the complementary functional at this order.

However, eq.(\ref{2nd.revisit.diff}) may be solved directly.
A similar situation occurs in potential theory
where one wishes to integrate an equation of the form
\begin{equation}
{\partial \phi \over \partial y_j}= g^j(y_k) \ ,
\end{equation}
which has the line-integral solution
\begin{equation}
\phi(y_k) = \int_C \sum_j dy^\prime_j \ g^j(y^\prime_l) \ ,
\end{equation}
where $C$ is an arbitrary contour whose upper end point
is $y_k$ and whose lower endpoint is arbitrary.
If the two endpoints are fixed, the solution is
independent of the choice of contour $C$
provided that $\sum g^jdy_j$ is exact.
By analogy, if $\int d^3x\;d\chi(x)\;{\cal R}^{(2)}$ is exact,
the solution of eq.(\ref{2nd.revisit.diff}) is
\begin{equation}
{\cal S}^{(2)}=-\int_C  \int d^3x \ d\chi'(x) \
{\cal R}^{(2)}\left[\chi'(x), f_{ij}(x)\right]
 \ .
\label{direct.integral.conf.2nd}
\end{equation}
The upper limit of integration
is $\chi(x)$ and the lower limit $\chi_0(x)$ is
an independent and arbitrary function.
The complementary functional is set to zero by choosing $\chi_0=0$.
We choose the simplest path for the line integral, that of a
straight line in superspace:
\begin{equation}
\chi'(x)=s\chi(x),\qquad d\chi'(x)=\chi(x)ds, \qquad 0\leq s\leq 1 .
\label{lineintegral}
\end{equation}
The real parameter $s$ is analogous to the Tomonaga-Schwinger
proper-time parameter
\cite{Tomo.Schw}.
The  integration over $s$,
$\int_0^1 ds \ s^{2/3} = 3/5$
is trivial leading to
\begin{equation}
{\cal S}^{(2)}=-\int d^3x \
{3\over 5}\chi\ {\cal R}^{(2)}\left[\chi, f_{ij}\right]
= \int d^3x f^{1/2} \ {3\over 5}\chi \left( {1\over2}\Omega \tilde R
- f^{ij}\chi_{,i}\chi_{,j} {\partial\ \over\partial\chi}(H\Omega)
- 2 f^{ij}\chi_{;ij} (H\Omega)\right)\ ,
\end{equation}
which, after an integration by parts, gives the same result as
before, eq.(\ref{second.order.in.f}).

\subsubsection{The fourth order}

We illustrate the conformal transformation method for the
fourth order.  We rewrite
${\cal R}^{(4)}$ in terms of the conformal metric $f_{ij}$ and
we find that, as before, we can integrate the equation directly:
it is exact.  The resulting generating functional written in terms
of $f_{ij}$ is
\begin{mathletters}
\label{Fourth.conf.dust}
\begin{equation}
{\cal S}^{(4)}=\int d^3x f^{1/2}
\left\{ l(\chi)\left(\tilde R^{ij}\tilde R_{ij} -{3\over 8}\tilde R^2\right)
+m(\chi)\tilde R^2
+n(\chi)\left(\tilde R^{ij}-{\tilde R\over 2}f^{ij}\right)\chi_{;i}\chi_{;j}
+r(\chi)\chi_{;i}^{\ \ ;i}\chi^{;j}\chi_{;j}
\right\}
\end{equation}
where
\begin{equation}
l(\chi)= - 2\int_0^\chi d\chi' \ {j^2(\chi')\over \Omega^3(\chi')} + L\ ,
\qquad
m(\chi)= M,\qquad
n(\chi)=-{j(\chi)\over \Omega^2(\chi)}\ ,\qquad
r(\chi)=-{1\over 4\Omega(\chi)}\ .
\end{equation}
\end{mathletters}The two arbitrary constants, $L$ and $M$, give
the additional complementary
functional.  Moreover, the complementary functional from the
second order gives
a contribution to this order through the dependence on $j$.  This is a
very succinct and tidy expression, which becomes more complicated when
expressed in terms
of the original 3-metric $\gamma_{ij}$, (using the conformal factor
(\ref{Omega.for.dust}))
\begin{eqnarray}
{\cal S}^{(4)}[\gamma_{ij}(x),\chi(x)] =&&-{3\over 14}\int d^3x\,
\gamma^{1/2}\, \left[ {9\over 25}\chi^3\left(R^{ij}R_{ij}-
{3\over 8}R^2\right) -{3\over 50}\chi R\chi_{|i}\chi^{|i}  \right.
\nonumber \\
 && \left.+ {3\over 25}\chi R^{ij}\chi_{|i}\chi_{|j} +
{19\over 50} \chi^{|i}\chi_{|i}\chi^{|j}_{\ \ |j}
-{1\over 3\chi} \chi^{|i}\chi^{|j}\chi_{|i}\chi_{|j}\right ]
\ ,
\label{fourthorderterms}
\end{eqnarray}
where we have neglected all the complementary functionals.

\subsubsection{The recursion relation}

If for simplicity we choose to ignore all the complementary
functionals at each order, we find that a simple recursion relation
may be derived which relates each order to the previous orders,
assuming that the remainder at each order is exact.
It is trivial but tedious to show that the remainders
${\cal R}^{(2)}$ and ${\cal R}^{(4)}$ are exact by functionally
differentiating ${\cal S}^{(2)}$,(\ref{second.order.in.f}),
and ${\cal S}^{(4)}$, (\ref{Fourth.conf.dust}).
A general proof of the exactness of the remainder ${\cal R}^{(2n)}$
and, as a result, the arbitrariness of the contour $C$ in the integral
\begin{equation}
{\cal S}^{(2n)}=-\int_C\int d^3x\ d\chi'(x)
{\cal R}^{(2n)}\left[\chi'(x), f_{ij}(x)\right]\ ,
\label{integral-for-dust}
\end{equation}
would rely on the fact that the Poisson bracket
$\{ {\cal H}(x), {\cal H}(y) \}$ yields the momentum constraint:
integrability  is assured provided gauge-invariance is maintained
at lower orders. (See also DeWitt \cite{DeWitt} and
Moncrief with Teitelboim \cite{M.and.T}.)

By induction, one can show that each term in
${\cal R}^{(2n)}$ on substitution of the line integral
(\ref{lineintegral}) has an $s$-dependence of $s^{2n/3}$.
Hence the integral of the HJ equation (\ref{integral-for-dust}) is
\begin{equation}
{\cal S}^{(2n)}=-\int d^3x\int_0^1  ds\  \chi(x) \
{\cal R}^{(2n)}[s\chi(x), f_{ij}(x)] =
-{3\over 2n+3}\int d^3x \ \chi(x)\  {\cal R}^{(2n)}[\chi(x), f_{ij}(x)]\ ,
\end{equation}
and we may rewrite this in terms of the 3-metric $\gamma_{ij}$
\begin{eqnarray}
{\cal S}^{(2n)}=&& -{3\over 2n+3}\int d^3x \ \chi(x)
{\cal R}^{(2n)}[\chi(x), \gamma_{ij}(x)] \nonumber \\
=&&-{3\over 2n+3}\int d^3\, x\left\{\gamma^{-1/2}\sum_{k=1}^{n-1}
\chi
{\delta{\cal S}^{(2k)}\over\delta\gamma_{ij}}
 {\delta{\cal S}^{(2n-2k)}\over\delta\gamma_{kl}}
 \left(2\gamma_{jk}\gamma_{li}- \gamma_{ij}\gamma_{kl}\right)
  \right.
\nonumber \\
 && \left. +\sum_{k=1}^n \chi
{{\scriptstyle 1\over2}\choose \scriptstyle p}
 \left(\gamma^{ij}\chi_{,i}\chi_{,j}\right)^k
{\delta{\cal S}^{(2n-2k)}\over\delta\chi}
+\chi{\cal V}^{(2n)}
\right\}\ .
\label{finalrec}
\end{eqnarray}
This expression has no reference to the conformal metric $f_{ij}$, and
hence is easy to use in practice.  In summary, to find the generating
functional at each order, one must functionally differentiate the
previous orders, and substitute in the above recursion relation.

\subsection{One Scalar Field}

To simplify the HJ equation (\ref{Htwon}) for a single scalar field we
use a similar conformal transformation to the dust field case
plus a change of variables for the scalar field.  Many of the
equations are similar.  Once again, $f_{ij}(x)$ is defined by
\begin{equation}
\gamma_{ij}(x) \equiv \Omega^2(u)f_{ij}(x)\ ,
\end{equation}
where the new variable $u(\phi)$ (which at the zeroth order is
a comoving, synchronous time variable) is defined through an integral
\begin{equation}
u=\int {d\phi\over -2{\partial H\over\partial \phi}}\ .
\label{def.of.u}
\end{equation}
The conformal factor satisfies the ordinary differential equation
\begin{equation}
{\partial \Omega\over\partial u}=H\Omega\ .
\end{equation}
For a single scalar field, the resulting HJ equation (\ref{Htwona})
has  a form similar to that of  dust:
\begin{equation}
\left. {\delta {\cal S}^{(2n)}\over\delta u}\right|_{f_{ij}} +
{\cal R}^{(2n)}
=0\ ,
\label{HJ.conf.for.scalar}
\end{equation}
which again reduces the system to a simple first order functional
differential equation with one independent variable.

${\cal R}^{(2n)}$ must be rewritten in terms of $f_{ij}$ and $u$.
At the second order we find that it is exact, and the HJ
equation (\ref{HJ.conf.for.scalar}) has the solution
\begin{mathletters}
\label{2nd.order.in.f.u}
\begin{equation}
{\cal S}^{(2)}=\int d^3x \ f^{1/2}
 \left[ j(u)\tilde R + k(u)u^{;i}u_{;i}\right]\ ,
\end{equation}
where
\begin{equation}
j(u)=\int^u_0 {\Omega(u')\over 2}\;du' + D\ ,\qquad
k(u)=H\Omega\ .
\end{equation}
\end{mathletters}This expression is remarkably similar to the dust
field case (\ref{second.order.in.f}).  For the exponential
potential considered previously (\ref{s.f.potential}) with the
zeroth order solution (\ref{zero.scalar}),
$u$ and the conformal factor are defined by
\begin{equation}
u=p/H, \qquad  \Omega\equiv u^p \ .
\end{equation}
These results can be used to show that the
generating functional above (\ref{2nd.order.in.f.u}) agrees with
our original approach (\ref{second.scalar}).

The fourth order solution also has a very similar form to the dust case.
\begin{mathletters}
\label{fourth.scalar.in.f.u}
\begin{equation}
{\cal S}^{(4)}=\int d^3x\  f^{1/2}\left\{
l(u)\left(\tilde R^{ij}\tilde R_{ij} -{3\over 8}\tilde R^2\right)
+m(u)\tilde R^2
+n(u)\left(\tilde R^{ij}-{\tilde R\over 2}f^{ij}\right)u_{;i}u_{;j}
+r(u)u_{\ ;i}^{;i}u^{;j}u_{;j}
\right\}
\end{equation}
where
\begin{equation}
l(u)= - 2 \int^u_0 du' {j^2(u')\over \Omega^3(u')} + L \ ,\qquad
m(u)=-\int^u_0 du' {\left({\Omega\over 2}- jH\right)^2\over 8
\left({\partial H\over\partial \phi}(u')\right)^2 \Omega^3(u')} + M\ ,
\qquad
n(u)=-{j\over \Omega^2}\ ,\qquad
r(u)=-{1\over 4\Omega}\ .
\end{equation}
\end{mathletters}
The main difference between this solution and the dust case
(\ref{Fourth.conf.dust}) is that here the coefficient of the
$\tilde R^2$ term depends on $u$.
Expressed in terms of the 3-metric $\gamma_{ij}$ and $\phi$ this
becomes quite complicated.
\begin{eqnarray}
{\cal S}^{(4)}= 2{p\over p-3}\left(p\over 2(p+1)\right)^2 \int d^3x
\ \gamma^{1/2}
&& {1\over H^3(\phi)}\left\{
R^{kl}R_{kl}+\left(-{3\over 8}+{1\over 8p}\right)R^2 +
{1\over\sqrt{2p}}R\phi^{\ \ |i}_{|i}\right.
\nonumber \\
&&\left. +{3\over 4}\left(1-{1\over p}\right) R\phi^{|i}\phi_{|i}
+\left({3\over 2p} - 2\right)R^{ij}\phi_{|i}\phi_{|j}
+\phi^{\ \ |i}_{|i} \phi^{\ \ |j}_{|j} \right.
\nonumber \\
&&\left. +\left(-1+{3\over 4p}\right){1\over\sqrt{2p}}
\phi^{\ \ |i}_{|i}\phi^{|j}\phi_{|j}
+\left({5\over 8} -{7\over 8p} +
{3\over 8p^2} \right)\phi^{|i}\phi_{|i}\phi^{|j}\phi_{|j}
\right\}
\end{eqnarray}

The recursion relation for a scalar field may be calculated in a
similar way to the dust case.  The integral of the HJ equation
(\ref{HJ.conf.for.scalar}) may be written down
\begin{equation}
{\cal S}^{(2n)}=- \int_{u_0}^u \int d^3x\  du'(x) \
{\cal R}^{(2n)}[u'(x), f_{ij}(x)]
\ ,
\label{scalar-integral}
\end{equation}
where $u_0$ is a constant function which we can set to zero
(this sets the complementary functionals to zero).
Once again, as in the dust case
(Sec. {\uppercase\expandafter{\romannumeral 3}}.A.3), the exactness
 of the second and fourth order remainders can be shown explicitly.
The exactness of the higher orders would seem to follow from the Poisson
brackets of the constraints.
One can solve (\ref{scalar-integral}) using a line integral as before
\begin{equation}
u'(x)=su(x),\qquad du'(x)=u(x)ds, \qquad 0\leq s\leq 1 .
\label{lineintegral.scalar}
\end{equation}
By induction, one can show that each term of ${\cal R}^{(2n)}$,
on substitution of this line integral, has an $s$-dependence
$s^{(2n-2)+(3-2n)p}$, and hence
\begin{eqnarray}
{\cal S}^{(2n)}=&&
-{1\over (2n-1) + (3-2n)p}\int d^3x \;u(x)\;{\cal R}^{2n}[u, f_{ij}]
\nonumber \\
=&&-{1\over (2n-1) + (3-2n)p}\int d^3x {p\over H(\phi)}\left\{
\gamma^{-1/2}\sum_{p=1}^{n-1}
 {\delta{\cal S}^{(2p)}\over\delta\gamma_{ij}}
 {\delta{\cal S}^{(2n-2p)}\over\delta\gamma_{kl}}
 \left(2\gamma_{jk}\gamma_{li}- \gamma_{ij}\gamma_{kl}\right)
\right. \nonumber \\
&&\qquad\qquad\qquad \qquad \qquad\left. + \gamma^{-1/2}
\sum_{p=1}^{n-1}{1\over 2} {\delta{\cal S}^{(2p)} \over\delta\phi}
    {\delta{\cal S}^{(2n-2p)}\over\delta\phi}
  + {\cal V}^{(2n)}\right\}\ .
\end{eqnarray}
Again, this expression has no explicit reference to the conformal metric,
and is easy to use.

\subsection{Multiple Fields}

One can solve the Hamilton-Jacobi equations for multiple
fields by employing the method of characteristics.
The characteristic equations are
\begin{mathletters}
\label{characteristics}
\begin{eqnarray}
{\partial\gamma_{ij}\over\partial\chi}=&& 2H\gamma_{ij}
\nonumber \\
{\partial\phi\over\partial\chi}=&&
-2{\partial{H}\over\partial\phi}
\end{eqnarray}
\end{mathletters}
with solution
\begin{mathletters}
\label{char.solution}
\begin{equation}
\gamma_{ij}=\Omega^{2}\left(\chi(x)\right)\,f_{ij}(x)\ ,
\label{defoff}
\end{equation}
\begin{equation}
\phi=h(\chi(x)+q(x)),
\label{phichar}
\end{equation}
\end{mathletters}
where the $f_{ij}(x)$ and the $q(x)$ are constants of integration
along a characteristic, and
\begin{equation}
{\partial{\Omega}\over\partial\chi} =H\Omega.
\label{choiceofomega}
\end{equation}
 In this way equation (\ref{Htwona}) has only one independent field
$\chi(x)$.
\begin{equation}
\left. {D{\cal S}^{(2n)}\over D\chi(x)}  \right|_{f_{ij},q}
=-{\cal R}^{(2n)}\ ,
\label{ham.2n}
\end{equation}
where $D$ denotes functional differentiation along a characteristic.
One of the characteristic equations (\ref{defoff}) can be seen
to define a conformal transformation.

\section{The evolution of the 3-metric}

Using an iteration scheme, we show how to solve the
evolution equation (\ref{evol.metric}) for the 3-metric.
We give explicit
solutions to fifth order for a homogeneous dust field.
A comparison is made with the Szekeres exact solution of
Einstein's equations. The Zel'dovich
approximation is briefly discussed. Firstly we find it
useful to define a restricted generating functional.

\subsection{The restricted generating functional}

At the end of the calculation, we are free to choose our time
hypersurface. For a single dust field, a
natural choice is a surface of uniform $\chi$, which
corresponds to comoving synchronous gauge ---
the evolution equation for $\chi$, (\ref{evol.dust}),
shows that the lapse $N=1$.
For now on, we will also choose the shift function $N^i=0$.

We define the restricted generating functional,
${\cal S}\left[\gamma_{ij}(x)|\chi\right]$, in which the dust field
is assumed to be homogeneous, by
\begin{equation}
{\cal S}\left[\gamma_{ij}(x)|\chi\right] \equiv
{\cal S}\left[\gamma_{ij}(x),\chi(x)=\chi\right].
\label{restr.funct}
\end{equation}
The rate of change of the restricted generating functional
${\cal S}[\gamma_{ij}(x)| \chi ]$ is
\begin{equation}
{ \partial { \cal S}[\gamma_{ij}(x)| \chi ] \over \partial \chi } =
\int d^3x \left [
{ \delta {\cal S}[\gamma_{ij}(x), \chi(x)]
\over \delta \chi(x) }
\right ]_{\chi(x)= \chi},
\end{equation}
in which case the Hamilton-Jacobi equation implies the
integrated Hamilton-Jacobi
equation for ${\cal S}[\gamma_{ij}(x)| \chi ]$
\begin{equation}
0=
{ \partial { \cal S}  \over \partial  \chi } +
\int d^3x \left \{  \gamma^{-1/2}
{ \delta { \cal S} \over \delta \gamma_{ij}(x) }
{ \delta { \cal S} \over \delta \gamma_{kl}(x) } \,
 \bigl[  2 \gamma_{jk}(x) \gamma_{il}(x)
- \gamma_{ij}(x) \gamma_{kl}(x) \bigr ]
-{ 1 \over 2}\gamma^{1/2} \, R  \right \},
\label{rr}
\end{equation}
and the momentum constraint reduces to
\begin{equation}
0 = -2 \left ( \gamma_{ik}
{ \delta { \cal S} \over \delta \gamma_{kj}(x) } \right )_{,j}
+ { \delta { \cal S} \over \delta \gamma_{lk}(x) } \gamma_{lk,i} \, .
\end{equation}
The restricted generating functional is much easier
to calculate in practice.  The gradient
expansion and the conformal transformation
trick work for eq.(\ref{rr}) as well. However,
integrability questions
are not important here because there is
only one equation to solve which always possesses solutions.

We can either compute the restricted functional using the
integrated HJ equation (\ref{rr}),
or using the full functional (\ref{fourthorderterms}) and setting
$\chi_{,i}=0$ at the end of the calculation.
Both methods give the same answer.
The restricted generating functional to fourth order is easily
seen from (\ref{fourthorderterms}).
\begin{equation}
{\cal S}^{(4)}\left[\gamma_{ij}(x)|\chi\right]=-\int d^3x\,
\gamma^{1/2}{27\over 350}\chi^3 \left(R^{ij}R_{ij} -{3\over 8}R^2\right).
\label{fourthdust}
\end{equation}
One can go even further and compute the sixth order functional
for a single homogeneous dust field:
\begin{eqnarray}
{\cal S}^{(6)}\left[\gamma_{ij}(x)|\chi\right]= \int d^3x\,\gamma^{1/2}\,
{ 54 \over 35 } \
\left\{-{49\over 64}R^3 +{29\over 8}RR^{ij}R_{ij}
\right.
\left.
-4R^{ij}R^k_{\ i}R_{jk}
+ R^{ij} R_{ij|k}{}^{|k}
-{3\over 8} R  R^{|k}{}_{|k} \right\}.
\label{sixth}
\end{eqnarray}

\subsection{The evolution of the 3-metric for dust}

One can use these results to compute the evolution equation
for the three-metric correct to higher order in spatial
gradients for a single dust field.  We adopt the iteration
scheme proposed in reference \cite{S.and.S}.
The evolution equation computed from the ADM action
(\ref{evol.metric}) on substitution of the spatial gradient
expansion of the generating functional, truncating at sixth order is
\begin{equation}
{ \partial \gamma_{ij} \over \partial \chi } = 2\gamma^{-1/2}
\left(2\gamma_{jk}\gamma_{il}-\gamma_{ij}\gamma_{kl}\right)\left(
{\delta{\cal S}^{(0)}\over\delta\gamma_{kl}}
+ {\delta{\cal S}^{(2)}\over\delta\gamma_{kl}}
+ {\delta{\cal S}^{(4)}\over\delta\gamma_{kl}}
+ {\delta{\cal S}^{(6)}\over\delta\gamma_{kl}}\right)\ ,
\label{evolutioneqn}
\end{equation}
where we have set the shift $N_i$ to be zero.
We choose our time coordinate to be $\chi$ which fixes the lapse
function $N=1$.

Retaining only the leading term on the right-hand side gives
the first order result for $\gamma^{(1)}(\chi, x)$,
eq.(\ref{first}). Substituting this expression into the right-hand
side, and retaining only third order terms, one finds
\begin{equation}
{ \partial \gamma_{ij}^{(3)} \over \partial \chi} =
2H \gamma_{ij}^{(3)} + { 3 \over 10 } \chi\,
\left (  \hat R k_{ij}- 4 \hat R_{ij} \right ).
\end{equation}
where $\hat R_{ij}\equiv R_{ij}( k_{lm}) $ and
$\hat R\equiv R( k_{lm} ) $ are the Ricci tensor and Ricci scalar,
respectively, of the seed metric $k_{lm}$.
The  third-order expression for the 3-metric is then:
\begin{equation}
\gamma^{(3)}_{ij}(\chi, x) =
\chi^{4/3} \, k_{ij}(x) +
{ 9 \over 20 } \chi^2 \,
\left [  \hat R(x) \,  k_{ij}(x) - 4 \hat R_{ij}(x)   \right ] .
\label{third}
\end{equation}
One more iteration leads to the fifth order expression
\begin{eqnarray}
\gamma^{(5)}_{ij} = &&\chi^{4/3} k_{ij}
+{9\over 20}\chi^2(\hat Rk_{ij}-4\hat R_{ij})
+ {81\over 350}\chi^{8/3}\left(\left\{ {89\over32}\hat R^2
+{5\over8}\hat R^{;k}{}_{;k} - 4\hat R^{kl}\hat R_{kl}\right\} k_{ij}
\right.
\nonumber \\
&& \left. -10 \hat R\hat R_{ij}+{5\over8}\hat R_{;ij}
+17\hat R_i^{\ n}\hat R_{jn} -{5\over2} \hat R_{ij;k}{}^{;k}\right)\ ,
\label{fourthordergamma}
\end{eqnarray}
where a semi-colon denotes covariant differentiation with respect to
$k_{ij}$.
A similar iteration scheme could be applied equally well to the
scalar field case.

\subsubsection{Comparison to an exact solution}

In this section we will compare our approximation scheme to
an exact solution of Einstein's equations --- the axisymmetric
specialization of the Szekeres
solution\cite{SZ} for irrotational dust, which has a line element
\begin{equation}
ds^2=-dt^2 + a^2(t)\left[dx^2+dy^2+(d(x^i)-c(z)a(t))^2dz^2\right],
\label{szekeres.line}
\end{equation}
where $d(x^i)$ has the form
\begin{equation}
d(x^i)=d_0(z)-{5\over 9}c(z)(x^2+y^2),
\label{dxi}
\end{equation}
and $d_0(z)$ and $c(z)$ are arbitrary functions of $z$.

We interpret our evolution equation (\ref{fourthordergamma}) as
evolving the seed metric $k_{ij}(x)$ which we choose to be
\begin{equation}
k_{ij}(x^i)=\hbox{diag}\;\left[1,1,d^2(x^i)\right],
\label{seed.metric}
\end{equation}
where $d(x^i)$ is given by (\ref{dxi}).
We can now evaluate all the terms on the right hand side of
(\ref{fourthordergamma}).  It is worthwhile using a
computer algebra program, particularly for the fourth order part.
Even for a simple metric such as (\ref{seed.metric}) the individual
terms such as $R_{ij;k}{}^{;k}$ are quite complicated, and so it is
extraordinary that the end result is so simple.  We find that
\begin{equation}
\gamma_{ij}^{(5)}=\chi^{4/3}\hbox{diag}\;
\left[1,1,(d(x^i)-c(z)\chi^{2/3})^2\right],
\label{final.metric}
\end{equation}
which is precisely the Szekeres solution (\ref{szekeres.line})!

We can use this to check that there is no seventh order contribution
to the Szekeres solution from our method by finding the solution
to the sixth order evolution equation.  We find that with the seed
metric (\ref{seed.metric}) there is
indeed no contribution to the metric from the sixth order,
i.e., $\gamma^{(7)}_{ij}=\gamma^{(5)}_{ij}$.

For situations of interest for cosmology, we do not
wish to assume such a special form for the seed metric.
For this case, the expansion can be improved quite significantly
if one notes
that the the determinant of the 3-metric must be non-negative. However,
one disturbing property
about the third order expression (\ref{third}) is that after
a sufficient amount of time,
the 3-metric may no longer be positive definate ----- not only does the
expansion break down, it gives nonsensical results.
One may remove this embarrassing problem by writing (\ref{third})
as a `square':
\begin{equation}
\tilde \gamma^{(3)}_{ij} =
\chi^{4/3}
\left \{ k_{il} + { 9 \over 40 } \chi^{2/3}
\left [ \hat R k_{il} - 4 \hat R_{il} \right ] \right \} \,
k^{lm} \,
\left \{ k_{jm} + { 9 \over 40 } \chi^{2/3}
\left [ \hat R k_{jm} - 4 \hat R_{jm} \right ] \right \}  \ .
\label{improved}
\end{equation}
Both eqs.(\ref{third}) and (\ref{improved}) agree to third order,
but the latter is the preferred form because it is non-negative.
This improved result reproduces the Zel'dovich approximation.
See ref. \cite{CPSS} and \cite{yamada} for a discussion of
higher order terms.

\section{Conclusions}

In this paper we have developed a systematic method for solving
the Hamilton-Jacobi equation for general relativity using a
spatial gradient expansion.  We derived a recursion
relation by which we can compute the generating
functional at each order by summing various
combinations of the functional derivatives of lower orders.
The key ingredients are a conformal transformation of the
3-metric and a line integral in superspace.

Our formalism provides some deep insights into the structure
of semiclassical superspace which now far exceeds
investigations in homogeneous models. Superspace
describes an ensemble of evolving universes, and
its complexity strains the imagination.
However, the gradient expansion allows one to
separate superspace into an infinite but discrete number of
manageable pieces which are relatively easy to understand.

Because the line integral in superspace (see
eq.(\ref{direct.integral.conf.2nd})) is independent of the
choice of contour (provided the endpoints are fixed), we now begin to
understand the  invariance of the generating functional
under different time-hypersurface choices. The exactness
of the remainder terms is highly nontrivial, and this
important property can essentially be traced to the fact
that the Poisson bracket of the Hamiltonian constraints,
$\{ {\cal H}(x), {\cal H}(y) \}$, yields the momentum constraint.
Integrability  is closely related to the gauge-invariance of the
theory.

The first few terms of the gradient expansion
are of considerable interest
for a dust-dominated Universe describing, for example, the
cold-dark-matter model. It is quite reassuring that the
fifth order 3-metric reproduces the exact Szekeres solution.
In addition, HJ theory provides an efficient and
practical means of calculating
higher terms in the Zel'dovich approximation \cite{CPSS},
\cite{yamada}.
However, we have not discussed the convergence of the  expansion.
For more general problems,
it seems likely that  one must employ techniques
that will improve the rate of convergence \cite{SSP}.

\acknowledgments

Both D.S.S. and J.P. were funded through the Science and
Engineering Research Council of the U.K..
D.S.S. also received
partial support from the Natural Sciences and Engineering
Research Council of Canada, and a Canadian Institute for
Theoretical Astrophysics National Fellowship
held in Edmonton.

\end{document}